\documentclass[10pt]{revtex4}
\usepackage{amssymb}
\usepackage{latexsym}
\usepackage{epsfig}                        
\usepackage{amsmath}
\begin{document}
\title{Thermodynamics of the apparent horizon in infrared modified Horava-Lifshitz gravity}
\author{Ahmad Sheykhi\footnote{asheykhi@shirazu.ac.ir}}
\address{ Physics Department and Biruni Observatory, College of
Sciences, Shiraz University, Shiraz 71454, Iran\\
  Research Institute for Astronomy and Astrophysics of Maragha
         (RIAAM), P.O. Box 55134-441, Maragha, Iran}

 \begin{abstract}
It is well known that by applying the first law of thermodynamics
to the apparent horizon of a Friedmann-Robertson-Walker universe,
one can derive the corresponding Friedmann equations in Einstein,
Gauss-Bonnet, and more general Lovelock gravity. Is this a generic
feature of any gravitational theory? Is the prescription
applicable to other gravities? In this paper we would like to
address the above questions by examining the same procedure for
Horava-Lifshitz gravity. We find that in Horava-Lifshitz gravity,
this approach does not work and we fail to reproduce a
corresponding Friedmann equation in this theory by applying the
first law of thermodynamics on the apparent horizon, together with
the appropriate expression for the entropy in Horava-Lifshitz
gravity. The reason for this failure seems to be due to the fact
that Horava-Lifshitz gravity is not diffeomorphism invariant, and
thus, the corresponding field equation cannot be derived from the
first law around horizon in the spacetime. Without this, it
implies that the specific gravitational theory is not consistent,
which shows an additional problematic feature of Horrava-Lifshitz
gravity. Nevertheless, if we still take the area formula of
geometric entropy and regard Horava-Lifshitz sector in the
Friedmann equation as an effective dark radiation, we are able to
extract the corresponding Friedmann equation from the first law of
thermodynamics.


\end{abstract}
 \maketitle

 \newpage
\section{Introduction\label{Intro}}
Inspired by Lifshitz theory in solid state physics, recently,
Horava proposed a field theory model for a UV complete theory of
gravity \cite{Horava}. This theory is a nonrelativistic
renormalizable theory of gravity and reduces to Einstein's general
relativity at large scales. The theory is usually referred to as
the Horava-Lifshitz (HL) theory. It also has manifest three-
dimensional spatial general covariance and time reparametrization
invariance. Various aspects of HL gravity have been investigated
in the literature.  A specific direction of the research is the
investigation of the thermodynamic properties of HL gravity. In
particular, black hole solutions in this gravity theory have been
attracted much attention. Here we review some thermodynamical
properties of HL gravity investigated previously, though our list
is not complete. Thermodynamics and stability of black holes in HL
gravity have been studied in \cite{Alex,My,cast,Wang}. It was
shown that HL black holes are thermodynamically stable in some
parameter space and an unstable phase also exists in other
parameter spaces \cite{CaiHL0}. In \cite{Jamil,Sama}, the validity
of the generalized second law of thermodynamics has been explored
in a universe governed by HL gravity. In \cite{CaiHL} the
relationship between the first law of thermodynamics and the
gravitational field equation of a static, spherically symmetric
black hole in HL gravity has been explored. It was shown that,
gravitational field equations of static, spherically symmetric
black holes in HL theory can be written as the first law of
thermodynamics on the black hole horizon \cite{CaiHL}. It was
argued that this approach can lead to extracting an expressions
for the entropy and mass of HL black holes which are consistent
with those obtained from other approaches \cite{CaiHL}. In the
cosmological setups, some attempts have been done to disclose the
connection between thermodynamics and gravitational field
equations of the Friedmann-Robertson-Walker (FRW) universe in HL
theory. For example, following the entropic interpretation of
gravity proposed by Verlinde \cite{Ver}, a modified Friedmann
equation in HL gravity was obtained in \cite{Shao}. Following
\cite{Shao}, the connection between the Debye model for the
entropic force scenario and the modified Friedmann equations in HL
gravity was also studied \cite{Bin}, although the results obtained
in \cite{Shao,Bin} seem to be  incorrect. The reason is that the
Friedman-like equation obtained in these papers from the entropic
force scenario is not the same as the one directly derived from
the field equations of HL gravity \cite{Alex,Jamil}. This
indicates that one cannot derive the Friedmann equation in HL
cosmology from the entropic gravity perspective. Other studies on
HL gravity have been carried out in \cite{otherHL}.

On the other side, it was shown that the gravitational field
equation of a static spherically symmetric spacetime in Einstein,
Gauss-Bonnet, and more general Lovelock gravity can be transformed
as the first law of thermodynamics \cite{Par}. The studies were
also extended to other gravity theories such as $f(R)$ gravity
\cite{Cai0} and scalar-tensor gravity \cite{Cai1}. In the
cosmological setup, it was shown that the differential form of the
Friedmann equation of FRW universe can be transformed to the first
law of thermodynamics on the apparent horizon \cite{Cai2,Cai3}.
The extension of this connection has also been carried out in the
braneworld cosmology \cite{Cai4,Shey1,Shey2}. The deep connection
between the gravitational equation describing the gravity in the
bulk and the first law of thermodynamics on the apparent horizon
reflects some deep ideas of holography.

Is the inverse procedure also always possible? (that is extracting
the general field equations from the first law of thermodynamics?)
Jacobson \cite{Jac} was the first who disclosed that the Einstein
field equation of general relativity can be derived from the
relation between the horizon area and entropy, together with the
Clausius relation $\delta Q=T\delta S$. For so called $f(R)$
gravity, Eling \textit{et al.} \cite{Elin}  argued that the
corresponding field equation describing gravity can be derived
from thermodynamics by using the procedure in \cite{Jac}, but a
treatment with nonequilibrium thermodynamics of spacetime is
needed. By using the entropy expression associated with the
horizon of the static spherically symmetric black hole solutions
in Einstein gravity, and replacing the horizon radius $r_{+}$ with
the apparent horizon radius, $\tilde{r}_{A}$, and taking the
ansatz for the temperature of the apparent horizon, it was shown
that the Friedmann equations of the FRW universe can be derived by
applying the first law of thermodynamics to the apparent horizon
for a FRW universe with any spatial curvature \cite{CaiKim}.
Employing the entropy relation of black holes to apparent horizon
in Gauss-Bonnet gravity and in the more general Lovelock gravity,
one also can get the corresponding Friedmann equations in these
theories \cite{CaiKim}. Also, starting from the first law of
thermodynamics, $dE=T_hdS_h+WdV$, at the apparent horizon of a FRW
universe, and assuming that the associated entropy with the
apparent horizon has a quantum corrected relation,
$S=\frac{A}{4G}-\alpha \ln \frac{A}{4G}+\beta \frac{4G}{A}$, one
is able to derive the modified Friedmann equations describing the
dynamics of the universe with any spatial curvature \cite{Shey3}.
These results indicate the holographic properties of the
gravitational field equations in a wide range of gravity theories.
For a recent review on the thermodynamical aspects of gravity see
\cite{Padrev}.

In the present work, we would like to address the question on the
connection between thermodynamics and HL gravity, by applying the
first law of thermodynamics on the apparent horizon of a FRW
universe and examine whether we can extract the corresponding
Friedmann equation in this gravity theory or not. Our strategy
here is to pick up the entropy expression associated with the
black hole horizon in HL gravity, assuming that the entropy
formula also holds for the apparent horizon of a FRW universe in
HL gravity. In other words, the apparent horizon has the same
expression for entropy but we replace the black hole horizon
radius $r_{+}$ by the apparent horizon radius $\tilde{r}_{A}$. We
find out that the resulting Friedmann equation from the first law
of thermodynamics differs from one obtained directly by varying
the action of HL gravity with respect to the FRW metric. This
shows an inconsistency in HL gravity that originates from the fact
that this theory is not diffeomorphism invariant, and thus, the
corresponding field equation cannot derive from the first law
around the horizon \cite{Pad2}.

This paper is structured as follows. In the next section we review
the IR modified HL theory and derive directly the corresponding
Friedman equation by varying the action. In Sec. III, we apply the
the first law of thermodynamics on the apparent horizon, together
with the appropriate expressions for the entropy and temperature,
and extract the Friedmann-like equation of the modified HL
cosmology. Only in the IR limit does the result of this section
coincide with the Friedmann equation obtained in Sec. II. In Sec.
IV, we assume the area law for the apparent horizon and derive an
effective Friedmann equation in modified HL gravity by applying
the first law of thermodynamics on the apparent horizon. We finish
our paper with a summary and discussion in Sec. V.
\section{Friedman Equation in IR modified HL cosmology\label{HL}}
In this section we first review the cosmological model which is
governed by HL gravity. The dynamical variables are the lapse and
shift functions, $N$ and $N_i$, respectively, and the spatial
metric $g_{ij}$. Using the Arnowitt-Deser-Misner (ADM) formalism,
the metric is parametrized as
\begin{equation}\label{ADM}
ds^2=-N^2 dt^2+g_{ij}(dx^i-N^idt)(dx^j-N^jdt).
\end{equation}
The Einstein-Hilbert action can be expressed as
\begin{equation}\label{EH}
I_{EH}=\frac{1}{16\pi G} \int{d^4x
\sqrt{g}N\left[(K_{ij}K^{ij}-K^2)+R-2\Lambda\right]},
\end{equation}
where $K_{ij}$ is the extrinsic curvature which takes the form
\begin{equation}\label{Kij}
K_{ij}=\frac{1}{2N}(\dot{g}_{ij}-\nabla _{i} N_{j}-\nabla _{j}
N_{i}),
\end{equation}
and a dot denotes a derivative with respect to $t$ and covariant
derivatives defined with respect to the spatial metric $g_{ij}$.
The action of HL gravity is given by \cite{Horava}
\begin{eqnarray}\label{SHL}
&&I_{SH}=\int{dtd^3x(\mathcal L_{0}+ \tilde {\mathcal L}_{1}+\mathcal L_{m})},\\
&&\mathcal L_{0}= \sqrt{g}N
\Big{\{}\frac{2}{\kappa^2}(K_{ij}K^{ij}-\lambda
K^2)+\frac{\kappa^2 \mu ^2 (\Lambda_W
R-3\Lambda_W^2)}{8(1-3\lambda)}\Big{\}}, \\
&&\tilde {\mathcal L}_{1}=\sqrt{g}N\Big{\{} \frac{\kappa^2
\mu^2(1-4\lambda)}{32(1-3\lambda)}R^2 -\frac{\kappa^2}{2 \omega^4
} \left(C_{ij}-\frac{\mu \omega^2}{2}R_{ij}\right)
\left(C^{ij}-\frac{\mu \omega^2}{2}R^{ij}\right) \Big{\}},
\end{eqnarray}
 where $\kappa^2$, $\lambda$, and $\omega$ are dimensionless
constant parameters, while $\mu$ and $\Lambda_W$ are constant
parameters with mass dimensions. Here $\mathcal L_{m}$ stands for
the Lagrangian of the matter field, $R$ and $R_{ij}$ are a
three-dimensional spatial Ricci scalar and Ricci tensor, and
$C_{ij}$ is the Cotton tensor defined as
\begin{equation}\label{Cij}
C^{ij}=\epsilon ^{ikl} \nabla _k \left(R^{j}_{\ l}-\frac{1}{4}
R\delta ^j_{l} \right)=\epsilon ^{ikl} \nabla _k R^j_{\
l}-\frac{1}{4}\epsilon ^{ikj} \partial _kR,
\end{equation}
where $\epsilon^{ikl}$ is the totally antisymmetric unit tensor.
It is worth mentioning that the IR vacuum of this theory is anti
de-Sitter (AdS) spacetimes. Hence, it is interesting to take a
limit of the theory, which may lead to a Minkowskian spacetime in
the IR sector. For this purpose, one may modify the theory by
introducing $\mu^4 R$ and then, taking the $\Lambda_W\rightarrow
0$ limit \cite{Alex}. This does not alter the UV properties of the
theory, but it changes the IR properties. That is, there exists a
Minkowski vacuum, instead of an AdS vacuum. We will now consider
the limit of this theory such that $\Lambda_W\rightarrow 0$. The
deformed action of the nonrelativistic renormalizable
gravitational theory is hence given by \cite{Alex}
\begin{eqnarray}\label{SHLdef}
I_{SH}=\int{dtd^3x\sqrt{g}N
\Big{\{}\frac{2}{\kappa^2}(K_{ij}K^{ij}-\lambda K^2)
-\frac{\kappa^2}{2 \omega^4 } C_{ij}C^{ij}-\frac{\mu^2
\kappa^2}{8}R_{ij} R^{ij} +\frac{\kappa^2 \mu}{2 \omega^2}
\epsilon ^{ijk} R_{il} \nabla _j R^l_{\ k}+ \frac{\kappa^2
\mu^2(1-4\lambda)}{32(1-3\lambda)}R^2+\mu^4 R \Big{\}}}.
\end{eqnarray}
In the IR limit, action (\ref{SHLdef}) can be written as the
standard Einstein-Hilbert action in the ADM formalism given in Eq.
(\ref{EH}), provided \cite{Alex}
\begin{equation}\label{cG}
\lambda=1, \  \   c^2=\frac{\kappa^2 \mu^4}{2}, \ \ \
G=\frac{\kappa^2 }{32 \pi c }.
\end{equation}
The constant $\omega$ is given by \cite{cast}
\begin{equation}\label{w}
\omega=\frac{8\mu^2 (3\lambda-1)}{\kappa^2}.
\end{equation}
Besides, for $\omega\rightarrow \infty$ (equivalently,
$\kappa^2\rightarrow 0$), action (\ref{SHLdef}) reduces to the
action of Einstein gravity \cite{cast,Wang} and hence we expect
all its solutions also recover to their respective ones in general
relativity.

We now consider a homogeneous and isotropic cosmological solution
to the theory (\ref{SHLdef}) with the standard FRW geometry
\begin{equation}\label{met1}
ds^2=-c^2dt^2+
a^2(t)\left[\frac{dr^2}{1-kr^2}+r^2(d\theta^2+\sin^2 \theta
d\Omega^2)\right].
\end{equation}
As usual, $k = -1, 0,+1$ corresponds to an open, flat, or closed
universe, respectively. Suppose that the energy-momentum tensor of
the total matter and energy in the universe has the form of a
perfect fluid $T_{\mu\nu}=pg_{\mu\nu}+(\rho+p)U_{\mu}U_{\nu}$
where $U^{\nu}$ denotes the four-velocity of the fluid and $\rho$
and $p$ are the total energy density and pressure, respectively.
The Friedmann equation, resulting from a variation of action
(\ref{SHLdef}) with respect to the FRW metric, turns out to be
\cite{Alex}
\begin{eqnarray}\label{FriHL1}
H^2=\frac{\kappa^2}{6(3\lambda-1)}\left(\rho-\frac{6k \mu^4}{a^2}
-\frac{3 k^2 \kappa^2 \mu^2}{8(3\lambda-1)a^4}\right),
\end{eqnarray}
where $H=\dot{a}/a$ is the Hubble parameter and we have imposed so
called projectability condition \cite{Char} and set $N=1$ and
$N^i=0$. The term proportional to $a^{-4}$ in (\ref{FriHL1}) is
the usual ``dark radiation" .  The energy conservation law
$\nabla_{\mu}T^{\mu \nu}=0$ leads to the continuity equation in
the form
\begin{equation}\label{Cont}
\dot{\rho}+3H(\rho+p)=0.
\end{equation}
For $k = 0$, there is no contribution from the higher order
derivative terms in the action. However, for $k\neq0$, the higher
derivative terms are significant for small volumes i.e., for small
$a$, and become insignificant for large $a$, where they agree with
general relativity. The standard Friedmann equation is recovered,
in units where $c = 1$,  provided we define
\begin{eqnarray}\label{Gc}
G_{\rm cosm}=\frac{\kappa^2}{16 \pi(3\lambda-1)},
\end{eqnarray}
and
\begin{eqnarray}\label{kap}
\frac{\kappa^2 \mu^4 }{3\lambda-1}=1,
\end{eqnarray}
where condition (\ref{kap}) also agrees for $\lambda=1=c$ with the
second relation in (\ref{cG}). Here $G_{\rm cosm}$ is the
``cosmological" Newton's constant. It is worth mentioning that in
theories such as HL, where the Lorentz invariance is broken, the
``gravitational" Newton's constant $G$ (that is, the one that is
present in the gravitational action) does not coincide with the
cosmological Newton's constant $G_{\rm cosm}$ (that is, the one
that is present in Friedmann equations), unless Lorentz invariance
is restored \cite{Jamil}. In the IR limit where $\lambda=1$, the
Lorentz invariance is restored, and hence $G_{\rm cosm} = G$.
Using the above identifications, as well as definition (\ref{w}),
the Friedmann equation (\ref{FriHL1}) can be rewritten as
\begin{eqnarray}\label{FriHL2}
H^2+\frac{k}{a^2}=\frac{8 \pi G_{\rm
cosm}}{3}\rho+\frac{k^2}{2\omega a^4}.
\end{eqnarray}
One can easily see that in the limit $\omega\rightarrow\infty$,
the dark radiation term vanishes and the standard Friedmann
equation is restored for $\lambda=1$ ($G_{\rm cosm} = G$), as
expected. On the other hand, one may also further rewrite the
Friedmann equation (\ref{FriHL2}) in the form
\begin{eqnarray}\label{FriHL3}
H^2+\frac{k}{a^2}=\frac{8 \pi G_{\rm cosm}}{3}\left(\rho+\rho_{\rm
rad}\right),
\end{eqnarray}
where in general $\rho=\rho_m+ \rho_{D}$ is the total energy
density of matter and dark energy and we have defined
\begin{equation}\label{rhorad}
\rho_{\rm rad}=\frac{3k^2}{16 \pi \omega a^4 G_{\rm cosm}}.
\end{equation}
Therefore, by regarding the dark radiation term which incorporates
the effect of HL gravity in the cosmological equation, we see that
the Friedmann equation of the modified HL cosmology can be written
completely in the form of standard cosmology. In such a case, one
does not take into account the richness of the HL gravity.  Note
that in the modified HL cosmology, $\rho_{\rm rad}$ cannot be
interpreted as the effective dark energy fluid, since at the late
time ($a\rightarrow\infty$) where the dark energy should be
dominated, it goes to zero. Besides, for a universe with spatial
curvature, it also has no contribution, which is not reasonable.
This is in contrast to HL cosmology in the presence of a
cosmological constant, where one can interpret the contribution
from $\Lambda$ and $\rho_{\rm rad}$ as an effective dark energy
\cite{Jamil,Sama}.

\section{Friedman Equation in HL cosmology from the First law \label{FIRST}}
In order to extract the cosmological equation governing the
evolution of the FRW universe in HL gravity, we need to have the
entropy expression of static spherically symmetric black holes in
this theory. It was argued that there is a deep connection between
the entropy expression and gravitational field equations
\cite{CaiKim,Shey3,CL}. Having the entropy expression at hand, one
can derive the corresponding cosmological equations in a wide
range of gravity theories including Einstein, Gauss Bonnet and
Lovelock gravity \cite{CaiKim}. The entropy expression depends on
the gravity theory and takes different forms for different gravity
theories. In the deformed HL gravity, the entropy associated with
the event horizon of the static spherically symmetric black holes
has the form \cite{Wang}
\begin{equation}\label{SHL}
S_h=\frac{A}{4G}+\frac{\pi}{\omega} \ln \frac{A}{G},
\end{equation}
where $A=4\pi r_{+}^2 $ is the area of the black hole horizon.
Throughout this paper, we set $k_{B}=c=\hbar=1$ for simplicity. As
one can see the entropy formula has a logarithmic term, which is a
characteristic of HL gravity theory \cite{Wang}. The parameter
$\omega$ can be regarded as a characteristic parameter in the
deformed HL gravity and the entropy relation will recover to the
well-known area law as $\omega\rightarrow \infty$. In general, the
entropy expression for HL black holes may have an additional
constant term $S_{0}$. However, as the HL parameter
$\omega\rightarrow \infty$, the HL gravity should be reduced to
the Einstein gravity; thus, we fix the constant term in the
entropy expression equal to zero.

We further assume the entropy expression (\ref{SHL}) is also valid
for the apparent horizon of the FRW universe in HL gravity.
Replacing the horizon radius $r_+$ with the apparent horizon
radius $\tilde{r}_{A}$, we have $A=4 \pi \tilde{r}_{A}^2$ for the
apparent horizon area in Eq. (\ref{SHL}). Now we take the
differential of the entropy (\ref{SHL}),
\begin{equation} \label{dS1}
dS=\frac{\partial S}{\partial A}dA=\frac{2 \pi
\tilde{r}_{A}}{G}\left(1+\frac{G}{\omega
\tilde{r}_{A}^2}\right)d\tilde{r}_{A}.
\end{equation}
The line element of the FRW universe can be written as
\begin{equation}
ds^2={h}_{a b}dx^{a} dx^{b}+\tilde{r}^2(d\theta^2+\sin^2\theta
d\phi^2),
\end{equation}
where  $x^0=t, x^1=r$, $\tilde{r}=a(t)r$, and  $h_{ab}$=diag $(-1,
a^2/(1-kr^2))$ represents the two-dimensional metric. Here $k$
specify the curvature of the spatial part of the metric. Solving
equation $h^{ab}\partial_{a}\tilde {r}\partial_{b}\tilde {r}=0$,
we obtain the dynamical apparent horizon radius of the FRW
universe as
\begin{equation}
\label{radius}
 \tilde{r}_A=\frac{1}{\sqrt{H^2+k/a^2}}.
\end{equation}
The temperature associated with the apparent horizon is defined as
$T_h = \kappa/2\pi$, where $
 \kappa =\frac{1}{2\sqrt{-h}}\partial_{a}\left(\sqrt{-h}h^{ab}\partial_{b}\tilde
 {r}\right)$ is the surface gravity.
It is easy to show that the surface gravity at the apparent
horizon of FRW universe can be written as \cite{Shey1}
\begin{equation}\label{surgrav}
 \kappa=-\frac{1}{\tilde
r_A}\left(1-\frac{\dot {\tilde r}_A}{2H\tilde r_A}\right).
\end{equation}
Since for $\dot {\tilde r}_A< 2H\tilde r_A$, we have $\kappa< 0$,
which leads to the negative temperature, one may, in general,
define the temperature on the apparent horizon as
$T_h=|\kappa|/2\pi$. In addition, since we associate a temperature
with the apparent horizon, one may expect that the apparent
horizon has a kind of Hawking radiation just like a black hole
event horizon. This issue was previously addressed \cite{cao} by
showing the connection between temperature on the apparent horizon
and the Hawking radiation. This study gives more solid physical
implication of the temperature associated with the apparent
horizon.

The next quantity we need to have is the work density. In our case
it can be calculated as \cite{Hay2}
 \begin{equation}\label{Work}
W=-\frac{1}{2} T^{\mu\nu}h_{\mu\nu}=\frac{1}{2}(\rho-p).
\end{equation}
Then, we suppose the first law of thermodynamics on the apparent
horizon of the universe in HL gravity holds and has the form
\begin{equation}\label{FL}
dE = T_h dS_h + WdV,
\end{equation}
where $S_{h}$ is the entropy associated with the apparent horizon
in HL cosmology which has the form Eq. (\ref{SHL}). The term $WdV$
in the first law comes from the fact that we have a volume change
for the total system enveloped by the apparent horizon. As a
result, the work term should be considered in the first law. For a
pure de Sitter space, $\rho=-p$, and the work term reduces to the
standard $-pdV$; thus, we obtain exactly the standard first law of
thermodynamics, $dE = TdS-pdV$.

Assume the total energy content of the universe inside a
three-sphere of radius $\tilde{r}_{A}$ is $E=\rho V$, where
$V=\frac{4\pi}{3}\tilde{r}_{A}^{3}$ is the volume enveloped by
three-dimensional sphere with the area of apparent horizon
$A=4\pi\tilde{r}_{A}^{2}$. Then we have
\begin{eqnarray} \label{dE1}
 dE &=&4\pi\tilde
 {r}_{A}^{2}\rho d\tilde {r}_{A}+\frac{4\pi}{3}\tilde{r}_{A}^{3}\dot{\rho}
 dt\nonumber \\
 &=&4\pi\tilde
 {r}_{A}^{2}\rho d\tilde {r}_{A}-4\pi H \tilde{r}_{A}^{3}(\rho+p) dt.
\end{eqnarray}
where we have used the continuity equation (\ref{Cont}) in the
last step. Substituting Eqs. (\ref{dS1}), (\ref{Work}) and
(\ref{dE1}) in the first law (\ref{FL}) and using the definition
of the temperature associated with the apparent horizon, we can
get the differential form of the Friedmann-like equation
\begin{equation} \label{Fried1}
\frac{1}{4\pi G}\frac{d\tilde {r}_{A}}{\tilde
{r}_{A}^3}\left(1+\frac{G}{\omega {\tilde {r}_{A}}^2}\right) = H
(\rho+p) dt.
\end{equation}
After using the continuity equation (\ref{Cont}), we reach
\begin{equation} \label{Fried2}
\frac{-2d\tilde {r}_{A}}{\tilde {r}_{A}^3}\left(1+\frac{G}{\omega
{\tilde {r}_{A}}^2}\right) = \frac{8\pi G}{3}d\rho.
\end{equation}
Integrating (\ref{Fried2}) yields
\begin{equation} \label{Fried3}
\frac{1}{{\tilde {r}_{A}}^2}+\frac{G}{2\omega {\tilde {r}_{A}}^4}=
\frac{8\pi G}{3}\rho,
\end{equation}
where an integration constant has been absorbed into the energy
density $\rho$. Substituting $\tilde {r}_{A}$ from
Eq.(\ref{radius}) we obtain a Friedmann-like equation of a FRW
universe in deformed HL gravity,
\begin{equation} \label{Fried4}
H^2+\frac{k}{a^2}+\frac{G}{2\omega
}\left(H^2+\frac{k}{a^2}\right)^2 = \frac{8\pi G}{3}\rho.
\end{equation}
As one can see the obtained Friedmann equation from the first law
of thermodynamics differs from the one obtained in (\ref{FriHL2})
from varying the action of the modified HL theory. This shows that
at least for the HL cosmology, the general prescription for
deriving the field equation from the first law of thermodynamics
does not work. This indicates a problematic feature of HL gravity.
In the IR limit of the theory where $\omega\rightarrow \infty$,
one recovers the standard Friedmann equation
\begin{equation} \label{Fried5}
H^2+\frac{k}{a^2} =\frac{8 \pi G}{3}\rho.
\end{equation}
Also for the late time cosmology where the scale factor $a$
becomes large, Eqs. (\ref{FriHL2}) and (\ref{Fried4}) coincide.
This is also consistent with the fact that the HL theory of
gravity reduces to Einstein's general relativity at large scales.
\section{An effective approach to HL thermodynamics\label{FIRST}}
In this section, we are going to use the effective approach and
try to reproduce the effective Friedmann equation of the deformed
HL cosmology given in (\ref{FriHL2}), by applying the first law of
thermodynamics on the apparent horizon. In the effective approach
all extra information of HL gravity can be absorbed in an
effective energy density and so, we consider that the universe
contains matter (and possible dark energy), plus this dark
radiation. Thus, the total energy density in the effective
approach can be written as $\rho_{\rm tot}=\rho+\rho_{\rm rad}$,
where $\rho= \rho_m+\rho_D$. The total energy density as well as
the dark radiation sector also satisfy the continuity equations
\begin{eqnarray}\label{contot}
&&\dot{\rho}_{\rm tot}+3H(\rho_{\rm tot}+p_{\rm tot})=0, \\
&&\dot{\rho}_{\rm rad}+3H(\rho_{\rm rad}+p_{\rm rad})=0.
\label{confrad}
\end{eqnarray}
Since the effective gravitational sector is now just the standard
general relativity, the entropy associated with the apparent
horizon obeys the well-known area formula,
\begin{equation}\label{Sef}
S_h=\frac{A}{4G_{\rm cosm}},
\end{equation}
where $A$ is the horizon area and $G_{\rm cosm}$ is the
cosmological Newton's constant given in (\ref{Gc}). Taking the
differential form of the above entropy, one finds
\begin{equation} \label{dSef1}
dS_h=\frac{2 \pi \tilde{r}_{A}}{G_{\rm cosm}}d\tilde{r}_{A}.
\end{equation}
Now we assume that the total effective energy content of the
universe is $E_{\rm eff}=\rho_{\rm tot} V$. Differentiating we get
\begin{eqnarray} \label{dEef1}
 dE_{\rm eff} &=&4\pi\tilde
 {r}_{A}^{2}\rho_{\rm tot} d\tilde {r}_{A}+\frac{4\pi}{3}\tilde{r}_{A}^{3}\dot{\rho}_{\rm tot}
 dt\nonumber \\
 &=&4\pi\tilde
 {r}_{A}^{2}\rho_{\rm tot} d\tilde {r}_{A}-4\pi H \tilde{r}_{A}^{3}(\rho_{\rm tot}+p_{\rm tot}) dt.
\end{eqnarray}
where we have used the continuity equation (\ref{contot}) in the
last step. Inserting Eqs. (\ref{Work}), (\ref{dSef1}) and
(\ref{dEef1}) in the first law of the form $dE_{\rm eff} = T_h
dS_h + WdV,$ and using the definition of the temperature
associated with the apparent horizon, we find the following
equation:
\begin{equation} \label{Fref1}
\frac{1}{4\pi G_{\rm cosm}}\frac{d\tilde {r}_{A}}{\tilde
{r}_{A}^3} = H (\rho_{\rm tot}+p_{\rm tot}) dt.
\end{equation}
After using the continuity equation (\ref{contot}), we reach
\begin{equation} \label{Fref2}
\frac{-2d\tilde {r}_{A}}{\tilde {r}_{A}^3}= \frac{8\pi G_{\rm
cosm}}{3}d\rho_{\rm tot}.
\end{equation}
Integrating and then substituting $\tilde {r}_{A}$ from
Eq.(\ref{radius}), we obtain the effective Friedmann equation of
the FRW universe in deformed HL gravity,
\begin{equation} \label{Fref4}
H^2+\frac{k}{a^2}= \frac{8\pi G_{\rm cosm}}{3}(\rho+\rho_{\rm
rad}).
\end{equation}
If we use the definition for $\rho_{\rm rad}$ given in Eq.
(\ref{rhorad}), we can further rewrite the above equation as
\begin{eqnarray}\label{Fref5}
H^2+\frac{k}{a^2}=\frac{8 \pi G_{\rm
cosm}}{3}\rho+\frac{k^2}{2\omega a^4},
\end{eqnarray}
which is exactly the result obtained in Eq. (\ref{FriHL2}). This
is an expected result, since in the effective approach, all extra
information of HL gravity has been absorbed in the energy sector
and the gravity sector is just standard Einstein's gravity.
However, in this case one does not take into account the richness
of the effects of gravity. The fact that the results of two
approaches coincide in the IR limit ($\omega\rightarrow\infty$)
can be easily understood, since in this case the correction term
in the entropy expression (\ref{SHL}) vanishes.
\section{Summary and discussion\label{Con}}
In summary, by applying the first law of  thermodynamics to the
apparent horizon of a FRW universe one can derive the
gravitational equations governing the dynamics of the universe in
a wide range of gravitational theories including Einstein
\cite{Jac}, Gauss-Bonnet, Lovelock \cite{CaiKim}, and $f(R)$
gravity \cite{Elin}. Can this prescription  be applied to other
gravitational theories?  In this paper we have shown that the
answer is not always positive. Having the entropy expression
associated with the horizon of spherically symmetric black holes
in infrared modified HL gravity, and applying it to the apparent
horizon, we failed to extract the corresponding Friedmann
equations in the modified HL gravity. This is the main result we
found in this paper, which originates from the fact that HL
gravity is not diffeomorphism invariant, and thus the
corresponding field equation cannot be derived from the first law
around horizon in the spacetime \cite{Pad2}. In order to justify
this result, here are some comments as follows.

(i) It was proved \cite{Pad2} that the field equations of any
theory of gravity which is diffeomorphism invariant must be
expressible as a thermodynamic identity, $TdS = dE$, around any
event in the spacetime. Also, in \cite {Pad2} it was shown that if
the theory is not diffeomorphism invariant, then the field
equation cannot be reexpressed as the first law.

(ii) The action of HL gravity is invariant under a restricted
class of diffeomorphism. The fundamental symmetry of HL theory is
the invariance under space-independent time reparametrization and
time-dependent spatial diffeomorphism  $ t\rightarrow t'(t), \  \
\vec{x}\rightarrow \vec{x}'(t,x)$ \cite{Kiri}. The time-dependent
spatial diffeomorphism allows an arbitrary change of spatial
coordinates on each constant time surface. However, the time
reparametrization here is not allowed to depend on spatial
coordinates. As a result, unlike general relativity, in HL gravity
the foliation of spacetime by constant time hypersurfaces is not
just a choice of coordinates but is a physical entity.

(iii) Also HL gravity treats space and time in a different way.
For the case of isolated black holes, the metric is time
independent and hence the spacetime is invariant under
infinitesimal diffeomorphism transformations. As a result,
according to the argument of \cite{Pad2}, the field equations of
the theory can be expressible as a thermodynamic identity, $TdS =
dE$ around the event horizon as it was shown in \cite{CaiHL}.

(iv) In the background of a FRW universe, the metric is time
dependent and hence the metric is not invariant under the
diffeomorphism transformation $ t\rightarrow t'(t), \  \
\vec{x}\rightarrow \vec{x}'(t,x)$. As a result, the field equation
cannot be expressed as the thermodynamics identity $TdS = dE$.
Without this, it implies that the specific gravitational theory
seems to be inconsistent and shows an additional problematic
feature of HL gravity. Our result in this paper confirms the
general discussion given in \cite{Pad2}.

It is important to note that if we still take the area formula of
geometric entropy and regard the HL sector in the Friedmann
equation as an effective dark radiation, we are able to derive the
corresponding Friedmann equation from the first law of
thermodynamics. Indeed, we have the freedom to choose two
approaches: the first is known as the robust approach and the
second is called the effective approach. In the robust approach,
we consider that the universe contains only matter (and possible
dark energy) and the effect of the gravitational sector of HL
gravity is incorporated through the modified entropy expression on
the horizon. In the effective approach, all extra information of
HL gravity can be absorbed in an effective energy density and so
we consider that the universe contains matter (and possible dark
energy), plus dark radiation. This approach is essentially same as
Einstein's gravity theory. We have shown that in the robust
approach, we failed to reproduce the corresponding Friedmann
equations in modified HL gravity. Nevertheless, in the effective
approach, we successfully derived the effective Friedmann equation
of modified HL gravity. Note that the results of the two
approaches coincide in the IR limit where
$\omega\rightarrow\infty$. This can be easily understood, since in
this limit HL gravity reduces to Einstein's gravity.

The results obtained in this paper, combined with those in
\cite{Jamil,Shao,Bin} indicate that HL gravity is not a consistent
gravitational theory from a thermodynamic perspective. Is it
possible to modify the HL theory in such a way that it can be
consistent with thermodynamics? This is quite an interesting
question, which deserves further investigation.

\acknowledgments{I thank the referee for very constructive
comments which helped me to improve the paper significantly. I am
also grateful to R. G. Cai, M. H. Dehghani and M. H. Vahidinia for
useful comments and helpful discussions. This work has been
supported financially by Research Institute for Astronomy and
Astrophysics of Maragha (RIAAM), Iran.}

\end{document}